\documentclass[prx,aps,groupedaddress,10pt,floatfix,twocolumn,amsmath,amssymb]{revtex4-1}

\usepackage{graphicx}
\usepackage{dcolumn}
\usepackage{bm}
\usepackage[hidelinks, bookmarks=false]{hyperref}
\usepackage{natbib}
\usepackage{newfloat}
\usepackage{units}
\usepackage{appendix}

\DeclareFloatingEnvironment[name={Figure S}]{suppfigure}

\makeatletter 
\def\tagform@#1{\maketag@@@{(A\ignorespaces#1\unskip\@@italiccorr)}}
\makeatother
 
\makeatletter
\makeatletter \renewcommand{\fnum@suppfigure}
{\figurename~S\thesuppfigure}
\makeatother
 
\makeatletter
\newcommand*{\balancecolsandclearpage}{%
  \close@column@grid
  \clearpage
  \twocolumngrid
}
\makeatother

\begin{document}

\title{Strong Optomechanical Squeezing of Light}
\author{T. P. Purdy}
\email{tpp@jila.colorado.edu}
\author{P.-L. Yu}
\author{R. W. Peterson}
\author{N. S. Kampel}
\author{C. A. Regal}
\affiliation{JILA, University of Colorado and National Institute of Standards and Technology,}
\affiliation{and Department of Physics, University of Colorado, Boulder, Colorado 80309, USA}

\date{\today}

\begin{abstract}
	We create squeezed light by exploiting the quantum nature of the mechanical interaction between laser light and a membrane mechanical resonator embedded in an optical cavity.  The radiation pressure shot noise (fluctuating optical force from quantum laser amplitude noise) induces resonator motion well above that of thermally driven motion.  This motion imprints a phase shift on the laser light, hence correlating the amplitude and phase noise, a consequence of which is optical squeezing.  We experimentally demonstrate strong and continuous optomechanical squeezing of $1.7\pm0.2$ dB below the shot noise level.  The peak level of squeezing measured near the mechanical resonance is well described by a model whose parameters are independently calibrated and that includes thermal motion of the membrane with no other classical noise sources.
\end{abstract}

\maketitle

	Interferometry is a ubiquitous method for sensitive displacement measurements.  In typical interferometry employing a coherent state, the amplitude and phase quantum fluctuations are both at the shot noise level.  Recently optomechanical systems have been developed that not only measure mechanical motion, but can also manipulate the motion with radiation pressure. For example, radiation forces have been used to cool mechanical resonators to near their quantum ground state~\cite{Teufel11, Chan11}.  With sufficiently strong radiation pressure, quantum fluctuations can become the dominant mechanical driving force, leading to correlations between the mechanical motion and the quantum fluctuations of the optical field~\cite{Purdy13}.  Such correlations can be used to suppress fluctuations on an interferometer's output optical field below the shot noise level~\cite{Fabre94,Mancini94}, at the expense of increasing fluctuations in an orthogonal quadrature.  This optomechanical method of manipulating the quantum fluctuations has historically been termed ponderomotive~\cite{Braginsky67} squeezing.

	The history of optical squeezing is intimately linked to quantum limited displacement sensing~\cite{Braginsky92}, owing to proposals to increase the displacement sensitivity of large scale gravitational wave observatories with squeezed light~\cite{Caves81, Unruh82, Jaekel90, Kimble01}.  Squeezed light was first produced using atomic sodium as a nonlinear medium~\cite{Slusher85}, and was soon followed with experiments employing optical fibers~\cite{Shelby86} and nonlinear crystals~\cite{Wu86}. Substantial squeezing has been achieved in modern experiments (up to 12.7 dB~\cite{Eberle10}), and enhanced sensitivity using squeezed light has been realized in gravitational wave detectors~\cite{LIGO11} and in biological measurements~\cite{Taylor13}.  Squeezed microwave fields, which have now been demonstrated with up to 10 dB of noise suppression~\cite{Beltran08}, are an important tool in quantum information processing with superconducting circuits.

\begin{figure*}[]
	\centering
		\includegraphics[width=\textwidth]{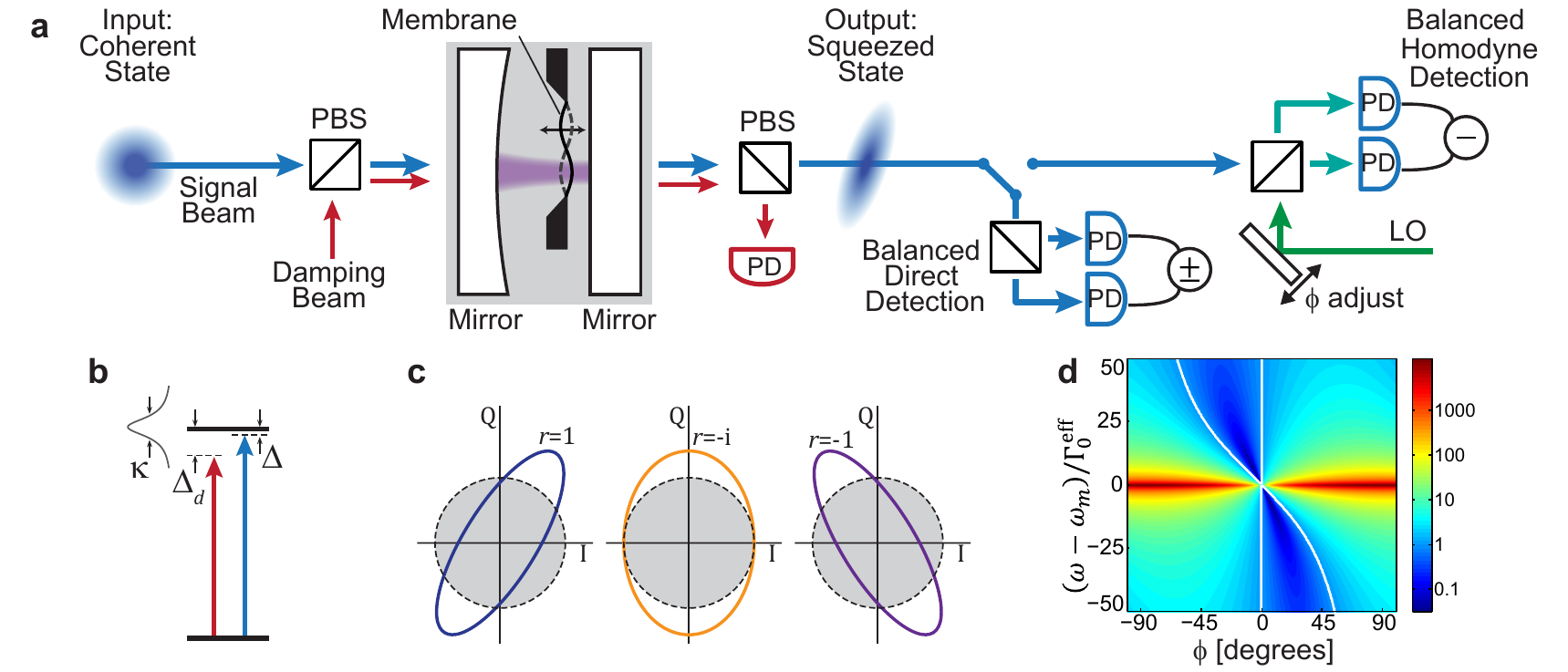}
	\caption{Experimental Diagram for Optomechanical Squeezing. (a) The signal beam (blue) enters the optomechanical cavity as a coherent state.  After the optomechanical interaction, a squeezed state emerges.  The signal beam is detected either with balanced direct photodetection or with balanced homodyne detection by mixing with an optical local oscillator (LO) (green).  A weaker damping beam (red) orthogonally polarized to the signal beam is also injected into the cavity.  The two beams are combined before the cavity and separated after the cavity with polarizing beamsplitters (PBS) and detected with photodetectors (PD). (b) Signal and damping beam detunings from the cavity resonance. (c) Representative $\delta X_{I}$-$\delta X_{Q}$ phase space distributions of the signal beam for real and imaginary values of $r$, the Kerr parameter.  The dashed circle represents the variance of the Gaussian noise distribution of the vacuum state.  Distributions inside the dashed circle represent squeezed states.  (d) Simulated signal beam quadrature spectrum for $\Delta=0$ in the idealized case of zero temperature and no optical loss.  Otherwise, Heisenberg Langevin simulation parameters are set to experimental values: $\kappa/2\pi=1.7$~MHz, $\Gamma_0^{\mathrm{eff}}/2\pi=2.6$~kHz, $\omega_m/2\pi=1.53$ MHz, and $\sqrt{\bar{N}} g_0/2\pi=350$~kHz. The spectrum is displayed on a logarithmic scale. The region between the white contours is squeezed.}
	\label{fig:Fig1}
\end{figure*}

	Early on, searches for ever-better squeezing materials led to suggestions that an optomechanical cavity, in which radiation pressure proportional to optical intensity changes the cavity length, could act as a low-noise Kerr nonlinear medium~\cite{Hilico92,Fabre94,Mancini94}, and hence could be a useful source of squeezed light~\cite{Corbitt06}. Further, a unique advantage of utilizing an optomechanical nonlinearity is that correlations induced by a mechanical object can be used to enhance displacement sensitivity for that same object~\cite{Kimble01,Heidmann97}.

	However, experimentally it has proven difficult to realize the substantial interplay between mechanical motion and quantum fluctuations of light required for ponderomotive squeezing.  Early on, radiation pressure induced optical non-linearity (bistability) was experimentally demonstrated in a cavity with a pendulum suspended end mirror~\cite{Dorsal83}.  More recently ponderomotively squeezed light at the few percent level has been demonstrated using a mechanical mode of an ultracold atomic gas inside an optical cavity~\cite{Brooks12}, and very recently using a  silicon micromechanical resonator~\cite{Safavi-Naeini13}.  The former experiment was limited by nonlinearities in the interaction and the latter by excess mechanical thermal motion.  Here, we observe ponderomotive squeezing at 1.7$\pm$0.2 dB below (32\% below) the shot noise level and optical amplification of quantum fluctuations by over $25~\unit{dB}$. The squeezing is realized on light transmitted through a Fabry-Perot optical cavity with an embedded mechanically compliant dielectric membrane.

	An optomechanical system can be thought of as an effective Kerr medium, and hence ponderomotive squeezing can be understood using many of the same ideas as typical nonlinear media.   However, in ponderomotive squeezing the finite mechanical response time defined by a complex mechanical susceptibility $\chi_m(\omega)$ plays an important role~\cite{Hilico92}.  We can illustrate the features of our expected ponderomotive squeezing by tracing the quantum fluctuations, $\delta X_{I}$ and $\delta X_{Q}$ in the optical amplitude and phase quadratures respectively, propagating through our optomechanical cavity (Fig.~1).  A large coherent state, referred to as the signal beam, consisting of nearly monochromatic radiation at a frequency $\omega_L$ and vacuum fluctuations at all other frequencies, enters the cavity from the left, and a vacuum state enters from the right.  We first consider the simplest case where the laser-cavity detuning, $\Delta$, is zero.  Because the membrane is located in a spatial gradient of the standing wave optical intensity, it is subject to an optical force from the shot noise intensity fluctuations, termed radiation pressure shot noise (RPSN).  The membrane responds to the RPSN drive with motion concentrated at frequencies near its mechanical resonances, i.e. weighted by  $\chi_m(\omega)$.  The mechanical motion of the dielectric membrane causes fluctuations in the cavity resonance frequency that are imprinted onto the optical phase quadrature, yielding $\delta X_{Q}(\omega)\rightarrow \delta X_{Q}(\omega)+r(\omega) \delta X_{I}(\omega)$, while $\delta X_{I}$ remains unchanged.  Here $r(\omega)$ is a dimensionless complex Kerr parameter proportional to the strength of the coupling, which depends upon a variety of parameters, including the mechanical response.  This Kerr-like self phase modulation correlates the amplitude and phase quadratures.  If $r$ is real, the correlations destructively interfere in some particular quadrature, leading to squeezing, as illustrated in Fig.~1(c).  If $r$ is purely imaginary the added phase fluctuations do not lead to squeezing.  The latter is the case when probing on the optical resonance and measuring at the mechanical resonance frequency, where the mechanical response is perfectly out of phase with the RPSN drive.
	
	Another consequence of the finite response time of the mechanical element (i.e. the imaginary component of $\chi_m(\omega)$) is that our system is directly coupled to the thermal bath.  Thermal motion of the membrane imprints excess noise onto the light, which is uncorrelated with the optical shot noise and hence can limit the level of squeezing.  To obtain significant squeezing near the mechanical resonance, the level of RPSN relative to the thermal force driving the membrane should be large~\cite{Purdy13}. For a beam with laser-cavity detuning near zero, this ratio is given by: $R=C/n_{\textrm{th}}\times(1+(2\omega_m/\kappa)^2)^{-1}$, where $C=4\bar{N} g^2/\kappa\Gamma_0$ is the optomechanical cooperativity, $\bar{N}$ is the average intracavity photon occupation, $g$ is the optomechanical coupling rate, $\Gamma_0$ is the mechanical dissipation rate, $\kappa$ is the cavity decay rate, $\omega_m$ is the mechanical resonance frequency, and $n_{\textrm{th}}$ is the thermal occupation of the mechanical state.
	
	Figure~\ref{fig:Fig1}(d) illustrates the basic features we expect to see when measuring the squeezing in homodyne detection as a function of quadrature angle $\phi$ and detuning with respect to the mechanical resonance.  To create this map we use the Heisenberg-Langevin model described in the Appendix that captures more of the complexity of our system.  In the diagram we can see that for pure intensity quadrature light we do not observe squeezing.  However, as one rotates towards the phase quadrature squeezing appears.  The lineshape of the squeezing is not symmetric about $\omega_m$ in our case, but instead follows a Fano-like lineshape.  The vacuum fluctuations directly reflected off the output mirror interfere with the quadrature-rotated, mechanical-resonance-modulated light exiting the cavity.  This diagram also illustrates the basic role of the mechanical susceptibility that weights the interaction between the membrane and the light.  Namely the magnitude of the squeezing ($S_\phi<1$) and the anti-squeezing ($S_\phi>1$) fall off on the scale of the mechanical linewidth.
	
	In our experiments we also operate with a finite laser-cavity detuning.   At a finite $\Delta$ an understanding of the spectrum of fluctuations must also take into account the $\Delta$ dependent quadrature rotation of the intracavity states relative to the input fields, $\phi_c=\tan^{-1}(2 \Delta/\kappa)$ (i.e. off-resonant phase fluctuations are partially converted into intracavity amplitude fluctuations and vice versa).  This quadrature rotation generates squeezing in the amplitude quadrature, which may be observed via direct photodetection in addition to homodyne detection. The spectral lineshape is also altered by the optomechanical damping and spring effects of the signal beam~\cite{Marquardt07} on the membrane's mechanical response (See Appendix).
	
\begin{figure*}[]
	\centering
		\includegraphics[width=16cm]{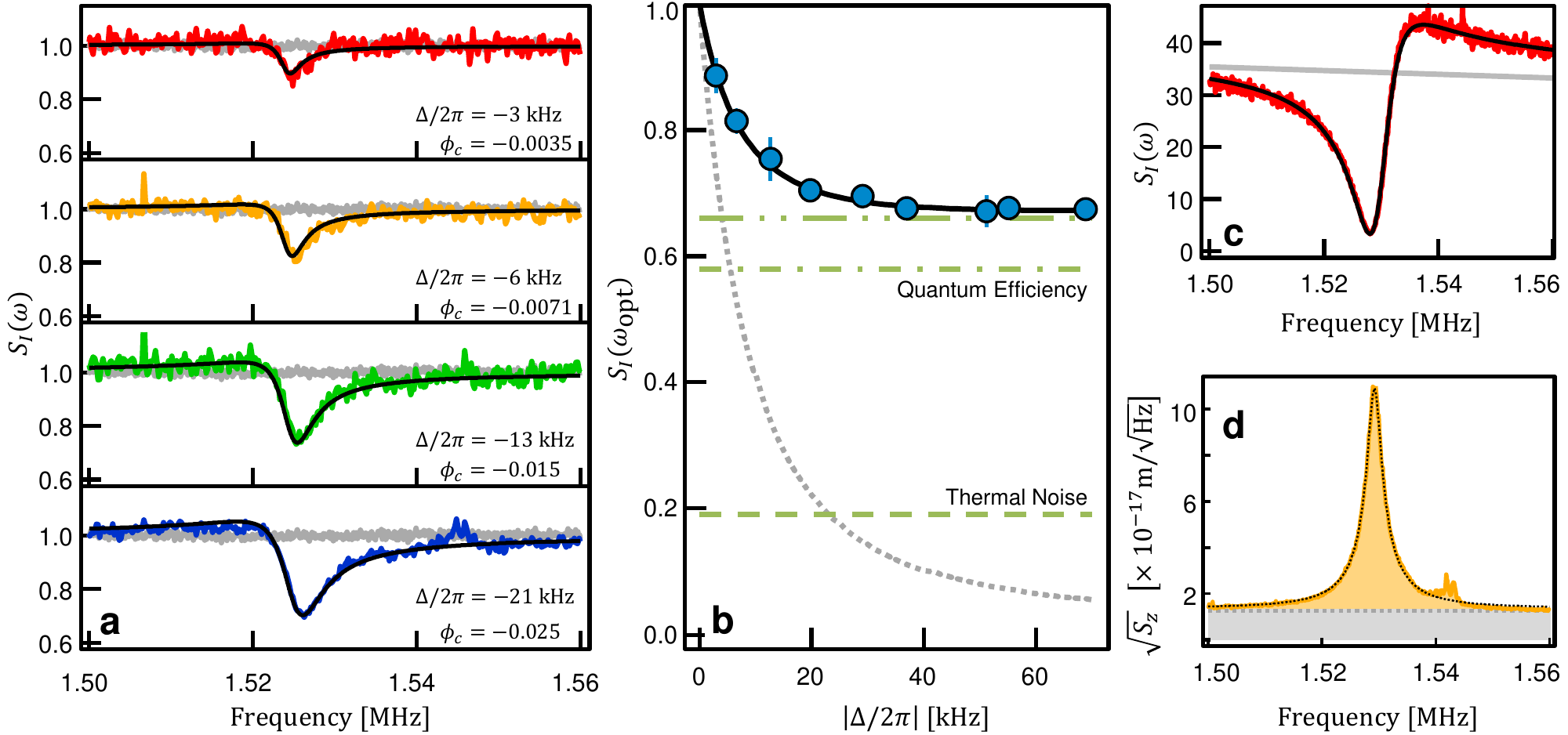}
	\caption{Quantum Intensity Noise Suppression.  (a) Directly detected optical intensity noise signal beam spectra for several signal beam detunings.  Also displayed are the measured shot noise level (gray) and the theoretical predictions (black).  A 200 Hz bandwidth is used.  The ratio of RPSN relative to thermal drive, $R$, is fixed at 5.1.  The damping beam provides $\Gamma_0^{\mathrm{eff}}/2\pi=2.7$ kHz and $\omega_m^{\mathrm{eff}}/2\pi=1.524$ MHz.  However the total mechanical damping rate and resonance frequency change with the signal beam detuning $\Delta$.  (b) The minimum value of $S_{I}$ for spectra as displayed in (a) (blue circles).  Statistical error bars indicate the standard deviation.  The frequency where the minimum occurs, $\omega_{\mathrm{opt}}$, shifts with $\Delta$ due to the optical spring effect.  Also displayed are the mechanical thermal noise floor for our current parameters $1/R$ (dashed green), limit set by finite detection efficiency $1-\epsilon$ (dot-dashed green), sum of thermal and detection efficiency limit (dot-dot-dashed green), and expected squeezing in the absence of optical loss and thermal motion (dotted gray).  (c) The directly detected optical intensity noise signal beam spectrum with intentionally added white, classical amplitude noise (red), theoretical prediction (black), and level of added amplitude noise (gray).  (d) The mechanical displacement spectrum inferred from the damping beam transmission spectrum (orange), and Lorentzian fit (dotted black). Detection noise floor is also shown (dotted gray).  One additional peak due to excess noise is visible in the bottom panel of (a) and in (d) at frequencies $\sim 1.545$ MHz due to a thermally occupied mechanical mode of the cavity support structure.}
	\label{fig:Fig2}
\end{figure*}


	Our optomechanical cavity (See Fig.~\ref{fig:Fig1} and Ref.~\cite{Purdy12}) consists of a 40 nm thick by 500 $\mu$m square silicon nitride membrane inside of a 3.54 mm long Fabry-Perot optical cavity~\cite{Thompson08}.  We work with the (2,2) drum-head mode of the membrane, with two antinodes along each transverse direction, yielding a mechanical resonance frequency of $\omega_m=2\pi\times 1.524$ MHz, and mechanical dissipation rate $\Gamma_0=2\pi\times0.22$ Hz.  The interaction Hamiltonian $\hbar g N z$, where $z$ is the operator of the effective mechanical coordinate and $N$ is the intracity photon number operator, is equivalent to that of a harmonically bound end mirror optomechanical cavity~\cite{Mancini94,Fabre94}.  In our system $g=2\pi\times33$ Hz, $\kappa=2\pi\times1.7$ MHz, and in a helium flow cryostat with a base temperature of 4.6 K, we achieve $R>5$, when operating with $\bar{N}\sim10^8$.  This ratio is much larger than achieved in previous work~\cite{Purdy13} mainly due to increased optomechanical coupling. 	In addition to our main signal beam, we inject another laser into the orthogonal polarization cavity mode.  This damping beam has a much weaker power than the signal beam, but is detuned by $\Delta_d\sim -\omega_m$ from the cavity resonance.  The damping allows us to avoid parametric instability and work with a mechanical mode with an effective mode temperature of less than $1~\unit{mK}$.  See Supplementary Materials for more details about experimental methods and calibrations.

	In our first set of experiments we use direct photodetection to measure the power spectrum of the amplitude quadrature, $S_{I}(\omega)$, which is normalized such that the detected shot noise is unity.  Figure~\ref{fig:Fig2}(a) shows $S_{I}$ for several values of $\Delta$, all at an average transmitted signal beam power of 110 $\mu$W corresponding to $\bar{N}=1.1\times10^8$ or $R=5.1$. A dip in noise below the shot noise level is visible in the vicinity of $\omega_m$, a clear signature of squeezed light.  The squeezing becomes more pronounced as $|\Delta|$ is increased because the maximally squeezed quadrature is rotated toward the amplitude quadrature.  The data show excellent agreement over most of its frequency range with a Heisenberg-Langevin model including quantum-noise-limited input optical fields, a thermally occupied mechanical bath coupled to the membrane, and no other classical noise sources (see Appendix). However, a small excess of classical noise is visible at the largest detuning, a few tens of kHz above the mechanical resonance.  Here, cavity frequency noise induced from a thermally occupied mechanical mode of the optomechanical cavity support structure~\cite{Purdy12} is increasingly converted in amplitude noise at larger $|\Delta|$.  All of the system parameters used to generate the theory curves of Fig.~\ref{fig:Fig2}(a) are independently measured, except $\Delta$ is calibrated, in part, using the displayed data.

	The shot noise level for the data of Fig.~\ref{fig:Fig2} is calibrated using balanced direct detection.  The transmitted signal is split into two equal power beams and directed onto a pair of nearly identical photodetectors.  Taking the sum of the detected signals is equivalent to single detector direct detection.  However, taking the difference of the detected signals removes classical and quantum correlations, up to the $20$ dB achieved common mode suppression.  The difference signal consists of only the uncorrelated shot noise level, and a small $\sim5\%$ contribution from the photodetector dark noise.  $S_{I}$ is computed by taking the ratio of the power spectra of the sum and difference signals, after subtracting the measured photodetector dark noise.

	The limits of the detected squeezing are illustrated in Fig.~\ref{fig:Fig2}(b) where the minimum measured value of $S_{I}$ is plotted as a function of $\Delta$.  The squeezing is limited somewhat by the thermal noise to RPSN ratio $1/R$.  The finite quantum efficiency of our detection system is the largest limit to the detected squeezing.  Including losses associated with the cavity $\epsilon_c=0.6$, propagation to the photodetector $\epsilon_p=0.8$, and photodetector conversion efficiency $\epsilon_d=0.87$, we estimate an overall quantum efficiency of $\epsilon=\epsilon_c\epsilon_p\epsilon_d=0.42$.

	For Fig.~\ref{fig:Fig2}(c) white classical intensity noise with an amplitude much greater than shot noise has been introduced onto the signal beam prior to entering the cavity.  While classical intensity noise is clearly suppressed as well, the lineshape of $S_{I}$ is qualitatively different from that of the quantum noise case.  In the classical noise case, the Fano asymmetry reverses due to the absence of coherent interference with phase and amplitude noise directly reflected from the output mirror, in contrast to the quantum noise case.  A Heisenberg-Langevin model incorporating the additional classical laser noise term agrees well with the measured data.  This symmetry difference between classical and quantum noise provides added confirmation that the spectra of Fig.~\ref{fig:Fig2}(a) truly arise from the manipulation of quantum noise.
	
	 Although the transmitted signal beam intensity spectrum is decidedly non-Lorentzian, the mechanical displacement still follows a simple Lorentzian form.  The damping beam transmitted intensity spectrum acts as a probe of mechanical motion uncomplicated by strong quantum correlations, because its intensity and thus RPSN effects on the membrane are small~\cite{Purdy13}.  The mechanical displacement spectrum derived from the damping beam (Fig.~\ref{fig:Fig2}(d)) shows the mechanics still retains a Lorentzian response to locally white force fluctuations.

\begin{figure}[]
	\centering
		\includegraphics[width=9 cm]{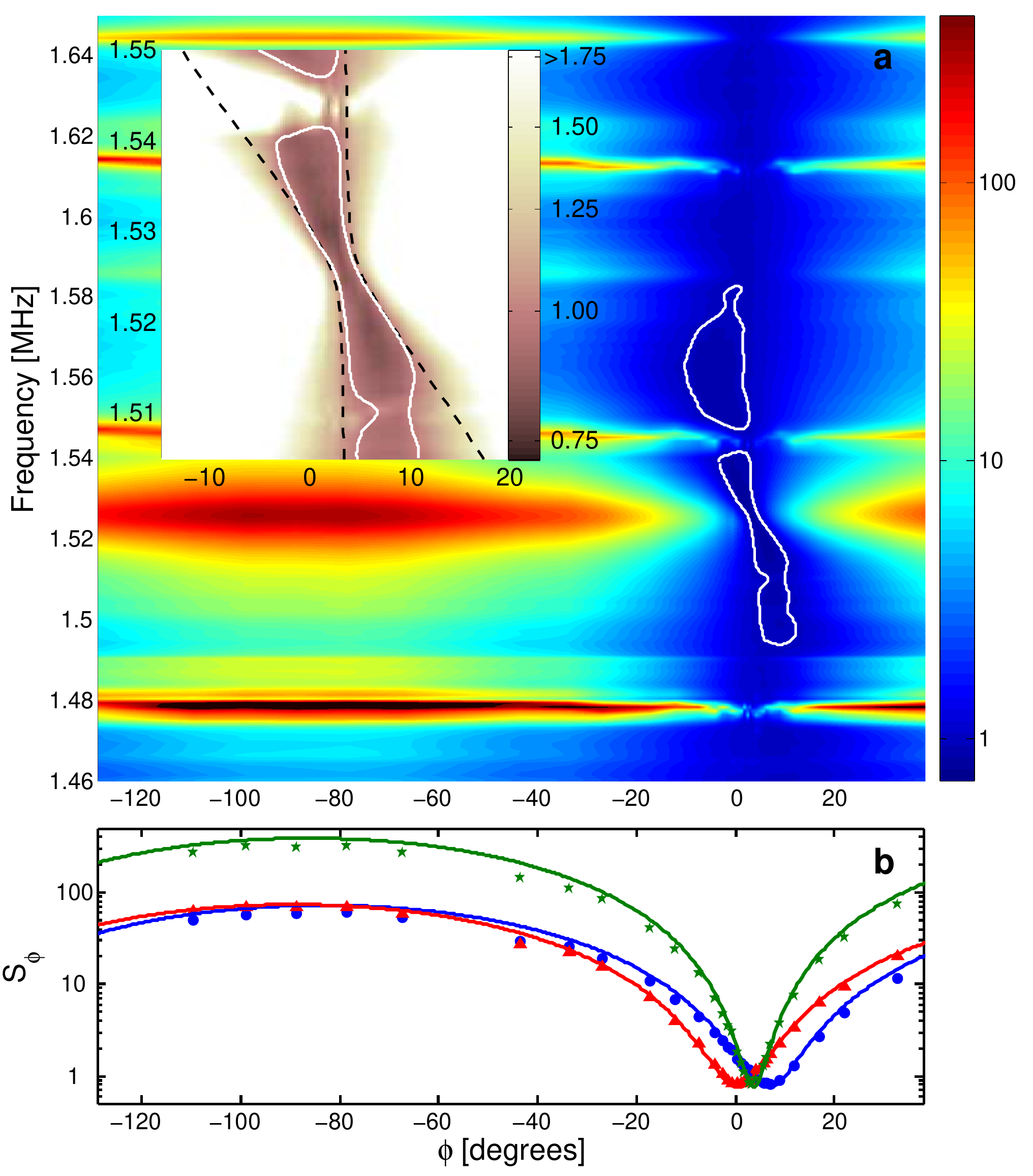}
	\caption{Optical Quadrature Spectrum.  (a) Color map of $S_{\phi}$.  The white contour is at the shot noise level, and the region inside this contour is squeezed.  Several additional noise peaks are evident at frequencies away from the mechanical resonance, due to motion of thermally occupied modes of the support structure.  The inset shows the squeezed region in more detail and also includes a theoretical prediction (dashed black) of the expected shot noise contour.  The experimental parameters are the same as in Fig.~\ref{fig:Fig2}, except $\Delta/2\pi=-42$~kHz.  (b) Cuts through quadrature phase at three different frequencies 1.517~MHz (blue circles), 1.526~MHz (green stars), and 1.535~MHz (red triangles), averaged over a 1 kHz bandwidth, and corresponding zero free parameter theoretical models (colored solid curves).}
	\label{fig:Fig3}
\end{figure}

	We next explore all quadratures of the transmitted signal beam with balanced homodyne detection.  We interfere the transmitted signal beam with an optical local oscillator whose phase is stabilized relative to the signal beam  (See Supplementary Materials for details).  In Fig.~\ref{fig:Fig3}(a) optical quadrature spectra, $S_{\phi}(\omega)$, over varying quadrature phase, $\phi$, are displayed.  $S_{\phi}$ is normalized such that the measured shot noise level is unity.  The phase $\phi=0$ corresponds the amplitude quadrature and gives information equivalent to that obtained in the direct detection discussed above.  (Note, for these measurements $\Delta=-2\pi\times 42$~kHz, allowing some squeezing to be rotated into the amplitude quadrature.)  Regions with noise spectral density below the shot noise level are visible over a bandwidth of $~\sim100$ kHz, and $\phi$ ranging over tens of degrees.  The range of observed squeezing is limited to a region smaller than predicted by theory because of the thermal motion of the cavity mirrors and support structure mentioned above.  The depth of the observed squeezing in homodyne detection is also lower than that observed in direct detection.  This is partially accounted for by imperfect overlap between the probe beam and homodyne local oscillator, contributing an additional effective optical loss, $\epsilon_m=0.8$.  Also, we operate in a regime where the homodyne local oscillator power is only a factor of less than 10 larger than the probe power, making the measurement slightly susceptible to the noise of the local oscillator (see Appendix).

	Homodyne detection also allows us to quantify the coherent amplification of optical quantum fluctuations in our measurement. In Fig.~\ref{fig:Fig3}(b) we compare our data to a theoretical calculation based upon a Heisenberg-Langevin model. The agreement between the model and the data allows us to interpret the large noise spectral density ($S_{\phi}\sim330$) near the phase quadrature at $\phi = \pm90^{\circ}$ as arising mainly from coherent amplification of quantum noise or so-called anti-squeezing.  This amplification persists despite the large imaginary component of the mechanical response, which has the potential to limit squeezing and add thermal noise.  Note, the measured spectral densities are far in excess of that required to satisfy the Heisenberg uncertainty limit, $\sqrt{S_{\phi}(\omega)}\sqrt{S_{\phi+\pi/2}(\omega)}\gg 1$. In the Supplementary Materials we present the parameters and configuration that would be required to realize a minimum uncertainty state that saturates the Heisenberg bound. 

	In conclusion, we have experimentally demonstrated that an optomechanical system well into the RPSN dominated regime is capable of creating squeezed light.  The $1.7~\unit{dB}$ strength of optomechanical squeezing we achieve is significantly larger than previous optomechanical realizations \cite{Brooks12,Safavi-Naeini13}. However, stronger squeezing has of course been realized with more developed techniques \cite{Eberle10}, and increasing efficiency and reducing thermal noise will be required to study the ultimate limits to deeply ponderomotively-squeezed light.  It will also be interesting to compare the passive squeezing achieved here to techniques that utilize optomechanically mediated quantum nondemolition measurements of the optical field and active feedback on the light \cite{Wiseman94,Mancini00}.

	This material is based upon work supported by the National Science Foundation under Grant Number 1125844, by the ONR young investigator program, and by the DARPA QuASAR program.  C.A.R. thanks the Clare Boothe Luce Foundation for support.


\setcounter{figure}{0} \renewcommand{\thefigure}{A.\arabic{figure}}

\section*{APPENDIX: CALCULATION OF OPTICAL SPECTRA} \label{sec: cal opt spec}
	In this Appendix we describe our solution to the Heisenberg-Langevin equations of motion for our optomechanical system.  We then compute the expected output optical quadrature spectrum, and the spectrum obtained from balanced homodyne detection and direct photodetection.
	
\subsection*{1. Heisenberg-Langevin Equations}
	We begin with the following Hamiltonian,  $H=H_0+H_{\kappa}+H_{\Gamma}$, where $H_0$ describes the intracavity coherent dynamics, $H_{\kappa}$ represents the coupling of the optical system to external fields, and $H_{\Gamma}$ represents the external thermal coupling to the mechanics~\cite{Marquardt07, Fabre94, Mancini94, Purdy13, Botter12}.
\begin{equation}
	H_0=\hbar \omega_m c^{\dag} c + \hbar \omega_c a^{\dag} a+\hbar G Z_{\mathrm{zp}} (c+c^{\dag}) a^{\dag} a
	\label{eq:Hamiltonian}
\end{equation}
where $\omega_m$ is the mechanical resonance frequency, $c$ $(c^{\dag})$ is the mechanical annihilation (creation) operator, $\omega_c$ is the optical resonance frequency, $a$ $(a^{\dag})$ is the optical intracavity annihilation (creation) operator,  $G$ is the optomechanical coupling constant, and $Z_{\mathrm{zp}}=\sqrt{\hbar/2 m \omega_m}$ is the mechanical zero point motion, with $m$ the mechanical resonator effective mass.  We define a single photon optomechanical coupling rate $g=G Z_{\mathrm{zp}}$, and a dimensionless mechanical displacement operator $z=\left(c+c^{\dag}\right)-\langle c+c^{\dag}\rangle$.   The Hamiltonian is linearized by assuming a large optical coherent state amplitude compared to the vacuum level, $a=(\bar{a}+d(t))e^{i\omega_L t}$, where $\omega_L$ is the optical drive frequency, $\bar{a}=\langle a \rangle$ is the intracavity coherent state amplitude, assumed to be real, and $d(t)$ is an operator containing the quantum and classical noise on the optical field. Terms of order $d^2$ are neglected. The linearized Hamiltonian that encapsulates the basic interaction is
\begin{align}
H_0=&\hbar \omega_m c^{\dag} c + \hbar \omega_c a^{\dag} a+\nonumber \\
       &\hbar G Z_{\mathrm{zp}} (c+c^{\dag}) \bar{a}^{*} \bar{a} + H_{BS} + H_{TMS} \label{eq:LinHamiltonian}\\
    H_{BS} =& \hbar G Z_{\mathrm{zp}} \left(\bar{a}^* c^{\dagger} d + \bar{a} c d^{\dagger}\right) \label{eq:BS_Ham} \\
	H_{TMS} = &\hbar G Z_{\mathrm{zp}} \left(\bar{a}^* c d + \bar{a} c^{\dagger} d^{\dagger}\right) \label{eq:PG_Ham}
\end{align}
The resulting linearized Hamiltonian contains both the beam-splitter ($H_{BS}$) and the two mode squeezing ($H_{TMS}$) Hamiltonians.

We solve the Heisenberg-Langevin equations of motion for this system in the frequency domain, using the Fourier transformation convention $f(\omega)\equiv\int^{\infty}_{-\infty}e^{\imath \omega t} f(t) dt$, $f^{\dag}(\omega)\equiv\int^{\infty}_{-\infty}e^{\imath \omega t} f^{\dag}(t) dt$, $\left(f^{\dag}(\omega)\right)^{\dag}=f(-\omega)$.  We assume thermally driven mechanical motion, with mechanical damping rate $\Gamma_0$ and initial thermal occupation $n_{\mathrm{th}}$.  We include the effects of the additional optical damping beam in an orthogonal cavity mode, by defining effective values for $\omega_m$, $\Gamma_0$, and $n_{\mathrm{th}}$ for the motion of the mechanical resonator in the presence of the optomechanical damping, spring, and cooling induced by the damping beam~\cite{Marquardt07}.  The optomechanical effects of the signal beam are intrinsic in the equations of motion.  The optical loss rate to the input port, output port, and internal loss are $\kappa_L$, $\kappa_R$, and $\kappa_{\mathrm{int}}$ respectively, yielding a total cavity damping rate of $\kappa=\kappa_L+\kappa_R+\kappa_{\mathrm{int}}$.  The external optical input fields consist of a coherent state, of frequency $\omega_L$, incident on the input port of the two-sided Fabry-Perot cavity, and vacuum incident on the output port.  An effective detuning of the input signal field from the average value of the optomechanically shifted cavity resonance is given by $\Delta$.  The optical output operator, $a_{\mathrm{out}}=(\bar{a}_{\mathrm{out}}+d_{\mathrm{out}}(t))e^{i \omega_L t}$, is computed using the cavity input-output relations $\bar{a}_{\mathrm{out}}=\sqrt{\kappa_R} \bar{a}$, $d_{\mathrm{out}}+d_{\mathrm{in}}= \sqrt{\kappa_R} d$, where $d_{\mathrm{in}}$ is the noise operator representing the vacuum field incident on the output port~\cite{Walls08}.

\subsection*{2. Optical Output Spectrum}
	The quadrature output operator is defined as $X_{\phi}(\omega)=a_{\mathrm{out}}(\omega)e^{i\phi}+a^{\dag}_{\mathrm{out}}(\omega)e^{-i\phi}$, where $\phi$ is the quadrature phase angle. Because we have assumed $\bar{a}$ to be real, the input-output relation indicate $\bar{a}_{\mathrm{out}}$ is also real, and $\phi=0$ ($\phi=90$) corresponds to the amplitude (phase) quadrature. The symmetrized power spectrum of the quadrature operator is $S_{XX}(\omega)$.

\begin{align}
	S_{XX}(\omega)=&\langle X_{\phi}(-\omega) X_{\phi}(\omega) \rangle_s \nonumber \\
	=&\frac{1}{2}\left(\langle X_{\phi}(-\omega) X_{\phi}(\omega) \rangle+\langle X_{\phi}(\omega) X_{\phi}(-\omega) \rangle\right) \nonumber \\
	=&A_{\zeta \zeta}(\omega)+A_{zz}(\omega)+A_{\zeta z}(\omega) 						
\end{align}
The spectrum consist of three terms. $A_{\zeta \zeta}$ is the shot noise on the output.  $A_{zz}$ represents the noise imprinted by the actual mechanical motion.  The cross term $A_{\zeta z}$ contains the correlations between shot noise and motion driven by radiation pressure from the shot noise. This term is responsible for any squeezing.

\begin{widetext}
\[
A_{zz}(\omega)=\kappa_R |\bar{a}|^2 g^2 \Big(|\chi_c(\omega)|^2+|\chi_c(-\omega)|^2-\chi_c(\omega)\chi_c(-\omega)e^{2 i \phi}-\chi^*_c(\omega)\chi^*_c(-\omega)e^{-2 i \phi}\Big) \langle z(-\omega) z(\omega)\rangle_s
\]
\begin{align*}
A_{\zeta z}(\omega)=i\sqrt{\kappa_R} \bar{a} g \Big(&
(-\chi_c(-\omega)e^{2 i \phi}+\chi^*_c(\omega))\langle z(-\omega) \zeta(\omega)\rangle_s +
(-\chi_c(\omega)e^{2 i \phi}+\chi^*_c(-\omega))\langle \zeta(-\omega) z(\omega) \rangle_s+\\
&(\chi^*_c(\omega)e^{-2 i \phi}-\chi_c(-\omega))\langle z(-\omega) \zeta^{\dag}(\omega) \rangle_s+
(\chi^*_c(-\omega)e^{-2 i \phi}-\chi_c(\omega))\langle \zeta^{\dag}(-\omega) z(\omega) \rangle_s
\Big)
\end{align*}
\[
A_{\zeta \zeta}(\omega)= \langle \zeta^{\dag}(-\omega) \zeta(\omega)\rangle_s+\langle \zeta(-\omega) \zeta^{\dag}(\omega)\rangle_s=1
\]
The output shot noise operator is $\zeta(\omega)=\chi_c(\omega)\sqrt{\kappa_L\kappa_R}\xi_L(\omega)+\chi_c(\omega)\sqrt{\kappa_{\mathrm{int}}\kappa_R}\xi_{\mathrm{int}}(\omega)+(\chi_c(\omega)\kappa_R-1)\xi_R(\omega)$.  $\xi_L$, $\xi_R$, and $\xi_{\mathrm{int}}$ are the Langevin vacuum noise operators for the input port, output port, and internal loss of the cavity respectively.
\[
\langle z(-\omega) z(\omega)\rangle_s=\frac{1}{|\mathcal{N}(\omega)|^2}\left(\Gamma_0 \left(\frac{n_{\mathrm{th}}+1/2}{|\chi_m(\omega)|^2}+\frac{n_{\mathrm{th}}+1/2}{|\chi_m(-\omega)|^2}\right)+ 2 \omega_m^2 g^2 \kappa |\bar{a}|^2 \left(|\chi_c(\omega)|^2+|\chi_c(-\omega)|^2\right)\right)
\]
\[
\langle z(-\omega) \zeta(\omega)\rangle_s=\frac{-\omega_m \bar{a} g \sqrt{\kappa_R}}{\mathcal{N}(-\omega)}\chi_c(\omega);\quad\quad\langle  \zeta(-\omega) z(\omega)\rangle_s=\frac{-\omega_m \bar{a} g \sqrt{\kappa_R}}{\mathcal{N}(\omega)}\chi_c(-\omega)
\]
\[
\langle \zeta^{\dag}(-\omega) \zeta(\omega)\rangle_s=\langle \zeta(-\omega) \zeta^{\dag}(\omega)\rangle_s=\frac{1}{2}
\]
\end{widetext}

We have introduced the following notation.  The cavity susceptibility is $\chi_c(\omega)=(\kappa/2-i(\Delta+\omega))^{-1}$.  The mechanical susceptibility is $\chi_m(\omega)=(\Gamma_0/2-i(\omega-\omega_m))^{-1}$. The optomechanical damping and spring effects are encompassed in $\mathcal{N}(\omega)=(\chi_m(\omega)\chi_m^*(-\omega))^{-1}-i 2 \omega_m g^2 |\bar{a}|^2 (\chi_c(\omega)-\chi_c^*(-\omega))$.  We also assume the mechanical thermal and optical vacuum baths are uncorrelated at different times $\left<\xi(-\omega')\xi(\omega)\right>=\delta(\omega-\omega')$, for noise operator $\xi$, and we assume integration over $\omega'$ for physically relevant quantities.

\begin{figure}[ht]
	\centering
		\includegraphics[scale=.8]{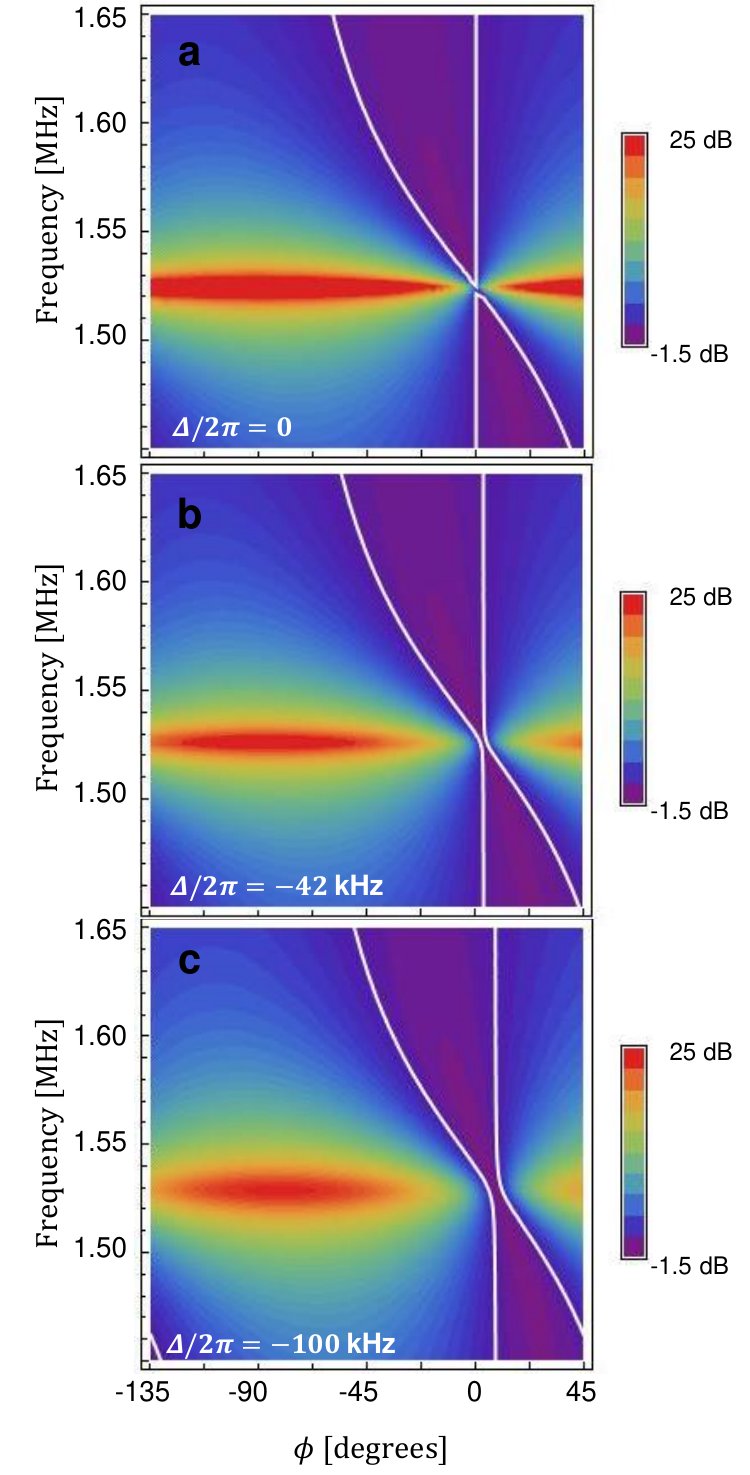}
	\caption{Calculated Homodyne Spectrum for finite signal beam detuning.  Homodyne transmission spectra, $S_{\phi}$, are calculated for three signal beam-cavity detunings: (a) $\Delta/2\pi=0$, (b) $\Delta/2\pi=-42$~kHz, (c) $\Delta/2\pi=-100$~kHz.  The other parameters used are $g/2\pi=33$~Hz, $m=6.75\times 10^{-12}$~kg, $\omega_m^{\mathrm{eff}}/2\pi=1.5243$~MHz, $\Gamma_0^{\mathrm{eff}}/2\pi=2560$~Hz, $T^{\mathrm{eff}}=3.8 \times 10^{-4}$~K, $\epsilon_{\mathrm{ext}}=0.55$, $\kappa/2\pi=1.7$~MHz, $\kappa_R=0.6\kappa$, $\bar{N}=1.1\times10^8$, $\epsilon_{\mathrm{ext}} |\bar{a}_{\mathrm{out}}|^2/|\bar{a}_{\mathrm{LO}}|^2=0.1$.  The parameters of panel (b) match the parameters of the measured spectrum presented in Fig.~\ref{fig:Fig3}.  With those parameters, the mechanical damping from the signal beam is $6~\unit{kHz}$. Calculated spectra are displayed on a logarithmic scale. The region between the white $0~\unit{dB}$ contours is squeezed.}
	\label{fig:FigA1}
\end{figure}

 Any loss in the optical detection system, including propagation losses between the cavity and detector, imperfect mode matching to the homodyne detector, or  finite photodetector conversion efficiency can be modeled by a single effective loss port with fractional loss $\epsilon_{\mathrm{ext}}$.  The loss port attenuates the signal reaching the detector $\bar{a}_{\mathrm{out}}\rightarrow\sqrt{\epsilon_{\mathrm{ext}}}\bar{a}_{\mathrm{out}}$, and injects vacuum noise leading an effective quadrature spectrum of $S_{XX}(\omega)\rightarrow \epsilon_{\mathrm{ext}} S_{XX}(\omega)+(1-\epsilon_{\mathrm{ext}})$.

	The homodyne detection consists of combining a strong optical local oscillator with the output from the cavity on a beam splitter.  Both outputs of the beam splitter are recorded on photodetectors and the two photocurrents are subtracted.  Assuming an equal splitting on the beam splitter, the subtracted photocurrent signal is proportional to $(\bar{a}_{\mathrm{out}}\bar{a}_{\mathrm{LO}}e^{i \phi}-\bar{a}_{\mathrm{out}}\bar{a}_{\mathrm{LO}}e^{-i\phi})+\bar{a}_{\mathrm{out}}(d_{\mathrm{LO}}+d^{\dag}_{\mathrm{LO}})+\bar{a}_{\mathrm{LO}}(d_{\mathrm{out}}e^{i \phi}+d^{\dag}_{\mathrm{out}}e^{-i \phi})$,  where the annihilation operator of the local oscillator is $a_{\mathrm{LO}}=(\bar{a}_{\mathrm{LO}}+d_{\mathrm{LO}}(t))e^{i \omega_L t+\phi}$, and we have neglected terms of order $d^2$.  The third terms is proportional to $X_{\phi}$.  The second term which represents the local oscillator vacuum noise beating against the coherent portion of the cavity output field is negligible when $\bar{a}_{\mathrm{LO}}\gg\bar{a}_{\mathrm{out}}$, and is typically ignored.  However, in the homodyne detection system described in the main text, we are limited to a local oscillator power which is less than 10 times larger than the signal beam power in order to ensure the photodetectors to remain in their linear range.  In this case, the local oscillator noise term must be included to quantitatively model the homodyne data.  The one-sided, symmetrized, shot noise normalized spectra, $S_{\phi}(\omega)$ in the main text are then given by:
\begin{align}
S_{\phi}(\omega)=\frac{2\left(|\bar{a}_{\mathrm{LO}}|^2(\epsilon_{\mathrm{ext}} S_{XX}(\omega)+(1-\epsilon_{\mathrm{ext}}))+\epsilon_{\mathrm{ext}}|\bar{a}_{\mathrm{out}}|^2\right)}{2\left(|\bar{a}_{\mathrm{LO}}|^2+\epsilon_{\mathrm{ext}}|\bar{a}_{\mathrm{out}}|^2\right)}
\label{eq:Sphi}
\end{align}
Using this full expression, one sees the level of perceived squeezing is reduced by the additional uncorrelated noise floor of the local oscillator.

	The one-sided, symmetrized, shot-noise-normalized, direct photodetection spectrum, discussed in the main text requires $S_{XX}(\omega)$ to be evaluated at $\phi=0$.
\begin{equation}
S_{I}(\omega)=\epsilon_{\mathrm{ext}} S_{XX}(\omega)|_{\phi=0}+(1-\epsilon_{\mathrm{ext}})
\end{equation}

	Several calculated spectra of $S_{\phi}$ in Fig.~\ref{fig:FigA1} illustrate the effects of the signal beam detuning, $\Delta$.  Three different values of $\Delta$ are displayed, and the other parameters are chosen to match the experimental data of the balanced homodyne experiment.  Two trends are evident as $\Delta$ is varied.  First, the entire spectrum is rotated by the cavity filtering by $\tan^{-1}(2\Delta/\kappa)$.  This is most evident by focusing on the white $0~\unit{dB}$ contour, which is shifted away from $\phi=0$ for increasing detuning.  Second, the optomechanical damping from the signal becomes significant for non-zero detuning, consequently broadening the features. This broadening is evident near the phase quadrature, which when $\Delta\sim0$ is proportional to the actual mechanical motion.  The area between the $0~\unit{dB}$ shot noise contours also becomes noticeably wider near the amplitude quadrature and near $\omega_m$ as $\Delta$ is increased.


\onecolumngrid
\clearpage

\begin{large}
	\begin{center}
	\begin{bf}
	Supplementary Information for: \\
	Strong Optomechanical Squeezing of Light
	\end{bf}
	\end{center}
\end{large}

\begin{center}
T. P. Purdy$^*$, P.-L. Yu, R. W. Peterson, N. S. Kampel, and C. A. Regal
\end{center}
\begin{center}
	\begin{it}
	JILA, University of Colorado and National Institute of Standards and Technology,\\
	and Department of Physics, University of Colorado, Boulder, Colorado 80309, USA\\[3\baselineskip]
	\end{it}
\end{center}

\twocolumngrid

\setcounter{figure}{0} \renewcommand{\thefigure}{S.\arabic{figure}}

\renewcommand{\bibnumfmt}[1]{[S#1]}
\renewcommand{\citenumfont}[1]{\textrm{S#1}}

\section{Experimental Setup}
\subsection*{Optomechanical Device}

\begin{suppfigure*}[t]
  \centering
  \includegraphics[width=11cm]{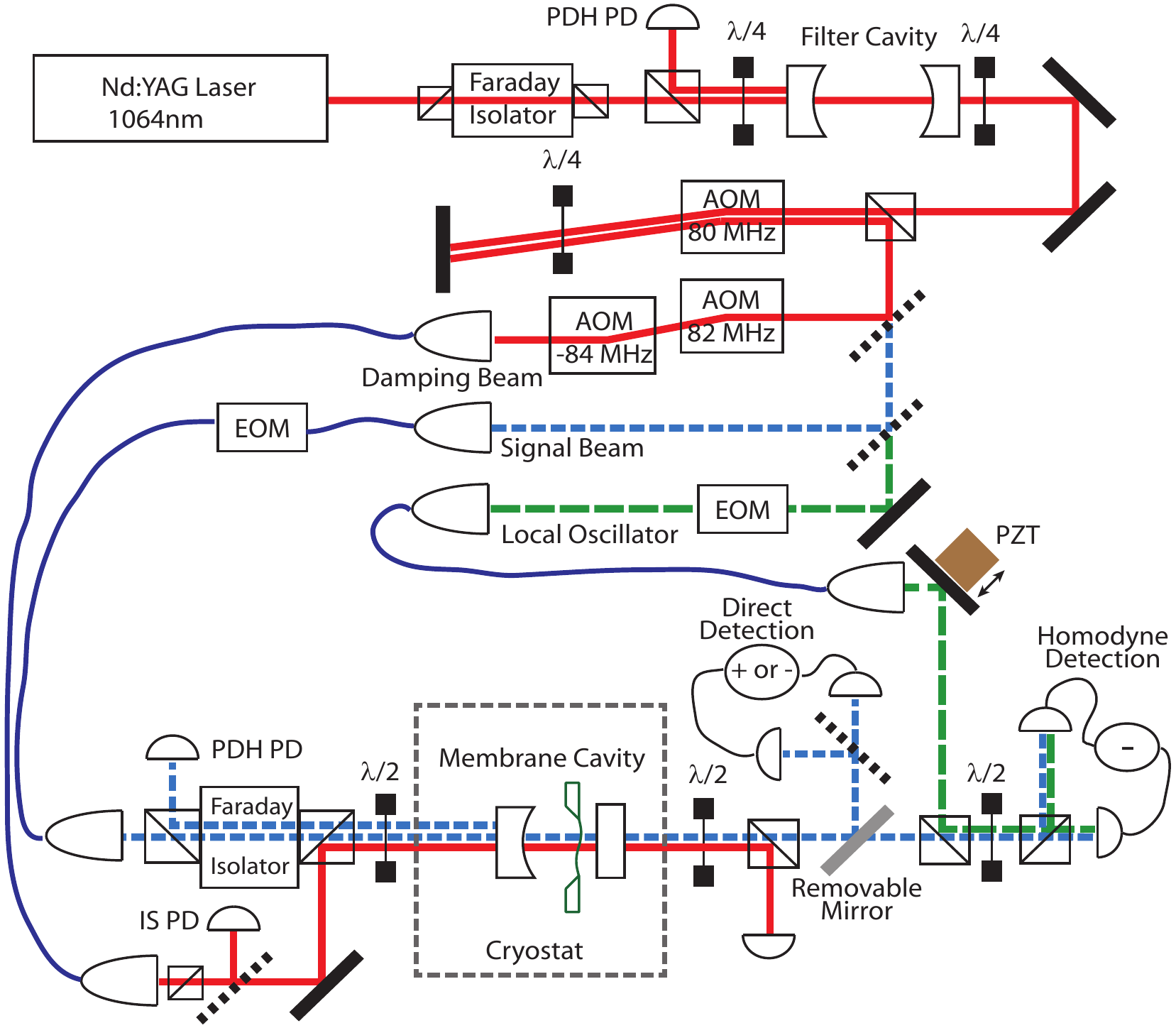}
\caption{Experimental Setup.   The signal beam (dotted blue), damping beam (solid red), and homodyne local oscillator (dashed green) are derived from a single passively filtered 1064 nm source.   Acousto-optical modulators (AOM) are used to shift the laser frequency.  An electro-optical modulator (EOM) is used to apply frequency sidebands for a Pound-Drever-Hall frequency lock on photodetectors (PHD PD).  The damping beam intensity is sampled on a photodetector (IS PD) used to actively stabilize its intensity.  A piezoelectric transducer (PZT) is used to stabilize the local oscillator path length.  Dashed black lines represent beam splitters, and boxes represent polarizing beam splitters.  $\lambda$/4 and $\lambda/2$ are quarter-wave and half-wave retarders respectively.  The transmitted signal is detected either with direct balanced photodetection or with balanced homodyne detection.}
\label{fig:figS5}
\end{suppfigure*}

	A detailed description of the construction and calibration of our membrane in a cryogenic cavity optomechanical system is described in Ref.~\cite{Purdy12S}. One key to the stable operation of our device is a monolithic design in which all elements are rigidly held in a cm-scale package.  Briefly, our cavity consists of one flat mirror and one 5 cm radius of curvature mirror.  For this work the mirrors are spaced 3.54 mm apart.  Both mirrors have a fractional intensity transmission of $1\times10^{-4}$.  The membrane consists of a 40 nm thick thin-film of high-stress stoichiometric silicon nitride with an index of refraction of approximately 2.  The membrane is a square of 500 $\mu$m on a side and is suspended in the center of a square single crystal silicon frame 5 mm on a side and 500 $\mu$m thick.  The membrane is held in the cavity about 900 $\mu$m from the flat mirror.  The curved end mirror and membrane are held by piezoactuators that allow about 500 nm of motion along the optical axis.  The end mirror actuator is employed to change the overall optical path length of the cavity.  The membrane actuator is used to position the membrane along the optical standing wave allowing operation at the maximum optical intensity gradient, where the linear optomechanical coupling is strongest.  Here, optomechanical coupling arises from two sources.  The dispersive optomechanical coupling changes the effective optical path length of the cavity as the optical intensity in the membrane changes.  An additional coupling occurs as the boundary conditions change for waves partially reflected from the membrane.  In this case there is an imbalance between the optical intensity on either side of the membrane leading to a radiation pressure coupling.  Using a one-dimensional matrix model~\cite{Wilson09S} for the three element optical cavity, along with the measured optical linewidth of $\kappa/2\pi=1.7$ MHz with the membrane at its nominal operating point, we estimate the input and output coupling and internal loss to be $\kappa_L=0.31 \kappa$, $\kappa_R=0.6\kappa$, and $\kappa_{\mathrm{int}}=0.09\kappa$ respectively.  We also measure a linear birefringence splitting of the polarization modes of the optomechanical cavity to be 300 kHz.  	

	The mechanical modes of the membrane are the transverse oscillation modes of a square, thin, high-tension drum.  We label each mode with two indices $(m,n)$ which indicate the number of antinodes of oscillation along each axis of the square.  The effective modal mass of any of these modes is $m=6.75\times 10^{-12}$ kg equal to 1/4 of the physical mass.  We focus on the $(2,2)$ mode of oscillation in this work, at a frequency of $\omega_m/2\pi=1.52$ MHz.  (Although we have observed squeezed light in the vicinity of several other low-order mechanical modes.)  The optical mode spot is centered near one of the antinodes of the (2,2) mechanical mode to obtain the maximum optomechanical coupling.
	
\subsection*{Optical Setup}	
	
	Our optical setup is similar to that described in Ref.~\cite{Purdy13S}, with the addition of balanced homodyne detection (Fig.~S1).  Our laser system consists of three beams derived from the same laser source, a low noise 1064 nm Mephisto laser from Innolight GmbH.  The laser is filtered by transmission through a 40 kHz linewidth Fabry-Perot cavity to remove classical noise in the measurement frequency band around 1.5 MHz.  The filtered beam is then double passed through an acousto-optical modulator (AOM), employed in the cavity frequency locking scheme.  The light is then split into three beams that act as the signal, damping, and homodyne local oscillator.   The damping beam passes through a pair of AOMs producing a beam shifted in frequency by a few MHz from the signal beam.  The signal beam passes through an electro-optical modulator that adds frequency sidebands at 30 MHz from the carrier, which are used to create a Pound-Drever-Hall error signal.  We employ a two branch feedback scheme to stabilize the laser-cavity detuning.  At low frequencies the end-mirror piezoactuator maintains the cavity to be nearly resonant with the signal beam.  At high frequencies, up to 100 kHz, the double passed AOM adjusts the laser frequency to follow fast deviations of the laser-cavity detuning.  The signal detuning $\Delta$ is set by adding a small electronic bias to the error signal.  The signal and damping beams are combined before the cavity on a polarizing beam splitter and split after the cavity with another polarizing beam splitter.  The absolute frequency difference between the signal and damping beam is set to be at least several hundred kHz from the measurement band near 1.5 MHz.  Hence, any interference from imperfect polarization separation does not affect the squeezing measurements.  The transmitted signal beam is then sent either to the direct photodetection system or the homodyne detection system.
	
	For the homodyne detection we combine the output signal beam and local oscillator on a polarizing beam splitter, rotate the polarization of the combined beams, and then split this light with another polarizing beam splitter.  This realizes an effective 50-50 beam splitter.  The light from each output port is directed onto a photodetector, and the photodetector signals are electronically subtracted.  This difference signal is then digitized at a sample rate of $2\times 10^7$ samples per second.  We compute the power spectrum of records of 25 ms in length.  Several thousand spectra are averaged for each measured phase quadrature in Fig.~3 of the main text, averaging over about 100 seconds worth of data.  We actively stabilize the measured quadrature phase by adjusting the homodyne beam path length via a piezoactuated mirror.  For $\phi$ near the phase quadrature $\phi=\pi/2$, we feedback to the average level of the photodetector difference.  For phases near the amplitude quadrature, $\phi=0$, an error signal is created by putting a small phase modulation on the homodyne beam with an electro-optical modulator.  The modulation sidebands at 5.6 MHz contain less than 1 percent of the total local oscillator power.  The photodetector difference signal is then mixed with an electronic local oscillator at the 5.6 MHz modulation frequency to create an error signal.  An electronic bias is added to this error signal to finely adjust the measured quadrature.  With this system, we are able to reduce fluctuations of the  homodyne phase to less than $1^{\circ}$ over the entire measurement range.  We are able identify the actual value of the phase to better than $2^{\circ}$ near the amplitude quadrature and better than $5^{\circ}$ near the phase quadrature.

\begin{suppfigure}[ht]
  \centering
   \includegraphics[width=9cm]{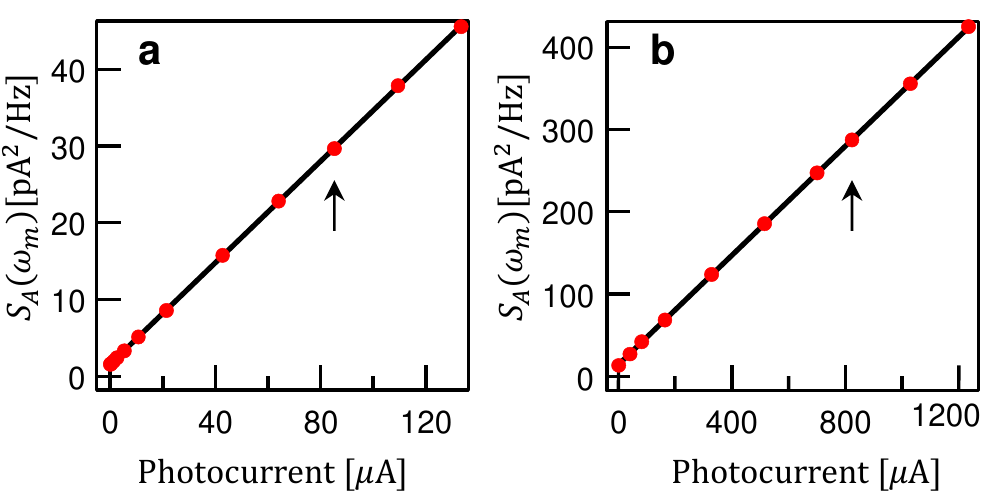}
    \caption{(a) Direct Detector Calibration. (b) Homodyne Detector Calibration.  Panel (a) (Panel (b)) displays the photocurrent noise power spectral density $S_A(\omega_m)$ as a function of detected photocurrent from the balanced detector used in direct (homodyne) detection (red circles). Each point is average over a bandwidth of 20 kHz centered around $\omega_m$.  The data are fit to lines (black).  The fitted slopes $3.305\pm0.005\times 10^{-19}$~A/Hz ($3.320\pm0.005\times 10^{-19}$~A/Hz) agree with expected value of $2 q_e=3.204\times10^{-19}$~A/Hz at the 3\% (4\%) level, indicating the detection is shot-noise limited and well calibrated for the direct (homodyne) detector. $q_e$ is the electron charge.  The arrows indicate the nominal operating point of the detector for the data taken in Fig. 2 (Fig.~3) of the main text for direct (homodyne) measurements.}
    \label{fig:figS4}
\end{suppfigure}

\begin{suppfigure}[ht]
  \centering
  \includegraphics[width=7cm]{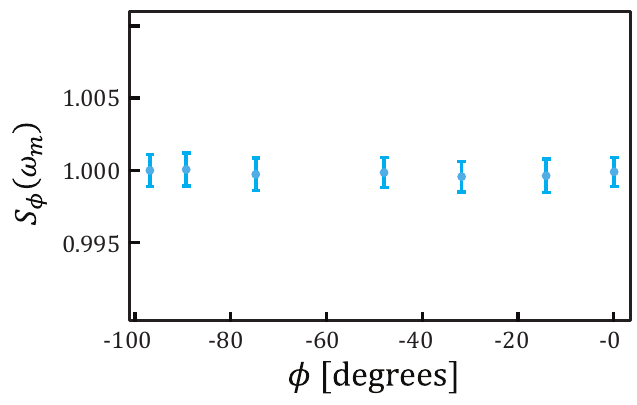}
\caption{Homodyne Calibration. $S_{\phi}$ from the homodyne balanced detector is plotted function of homodyne angles for a shot noise limited signal beam that does not pass through the optomechanical cavity. The blue circles are data and the error bars indicate the statistical standard deviation. The data are taken with a 832~$\mu$W local oscillator beam and a 110~$\mu$W signal beam (the same optical powers used in Fig.~3 of the main text).  Each point is average over a bandwidth of 50~kHz around $\omega_m$.}
\label{fig:figS5}
\end{suppfigure}

\section{Photodetector and Laser Noise Calibration}

\subsection*{Photodetector Linearity}
	The balanced detectors for both direct detection and homodyne detection consist of Hamamatsu G10899 series InGaAS PIN photodiodes with a measured quantum efficiency of 0.87.  We demonstrate the linearity of the detectors in Fig.~S2.  For the direct detector, we shine light that does not pass through the optomechanical cavity onto the detector.  For the homodyne detector we vary the local oscillator power with no signal beam present.  The detected noise shows a highly linear trend with detected optical power for both detectors, indicating shot noise limited detection, excellent linearity, and low photodetector dark noise. 	We also test the linearity over quadrature phase in Fig.~S3.  Here, we perform homodyne detection with a signal beam that does not pass through the optomechanical cavity.  We observe a flat, shot-noise-limited response over quadrature angle.

\subsection*{Classical Laser Intensity Noise and Shot Noise Calibration}	
	We measure that the intensity noise on the signal beam to be shot noise limited before passing through the optomechanical cavity by comparing the sum and difference of balanced photodetectors.  This test shows the amplitude quadrature of the signal be to shot noise limited to better than 0.1 dB.  The calibration of this measurement is limited by the slight mismatch of the amplifier gains of the summing and differencing electronics to the 0.1 dB level.  Further, our agreement between the measured laser noise levels and those expected from careful independent calibration of the photodetector response (see Fig~S2) is also at the 0.1 dB level.  This systematic uncertainty in the shot noise level between calibration techniques leads to the quoted uncertainty on the maximum level of squeezing obtained in the main text of $\pm0.2$ dB.

\subsection*{Laser Phase Noise}	
  We also independently assess the technical noise on the phase quadrature of the signal beam before it passes through the optomechanical cavity.  Because the signal beam and local oscillator are derived from the same laser source, common mode noise between the two, introduced from electronic noise on the common path AOM for instance, will be rejected by balanced homodyne.  However, this noise will become apparent when the signal beam is filtered by the optomechanical cavity, destroying the common mode rejection at frequencies in the measurement band. By passing the local oscillator through an additional filter cavity with a linewidth of $\sim 100$ kHz (before it interferes with a signal beam that does not pass through the cavity) we can remove the noise correlations and gain an independent measurement of the signal beam noise.   Using this technique, we estimate the phase noise on the signal beam to less than 1 dB above shot noise in the measurement band near the 1.52 MHz membrane resonance at an operating power of 110 $\mu$W. This phase noise does not effect the squeezing results obtained near the amplitude quadrature for parameters employed in the main text.  For $\Delta \ll \kappa$, the frequency to amplitude noise conversion by cavity quadrature rotation is small, i.e.~$\tan^{-1}(2\Delta/\kappa)$ is only a few degrees for the experimental parameters of the main text.  However, for operation at larger detuning, for instance $\Delta\sim\omega_m$, the excess phase noise could potentially pose a considerable limit to squeezing.  However, in our current experiment, the most significant source of laser-cavity frequency noise is the thermally induced mechanical motion of the mirror and membrane substrates~\cite{Zhao12S,Purdy12S}.

\begin{suppfigure}[ht]
	\centering
		\includegraphics[scale=1]{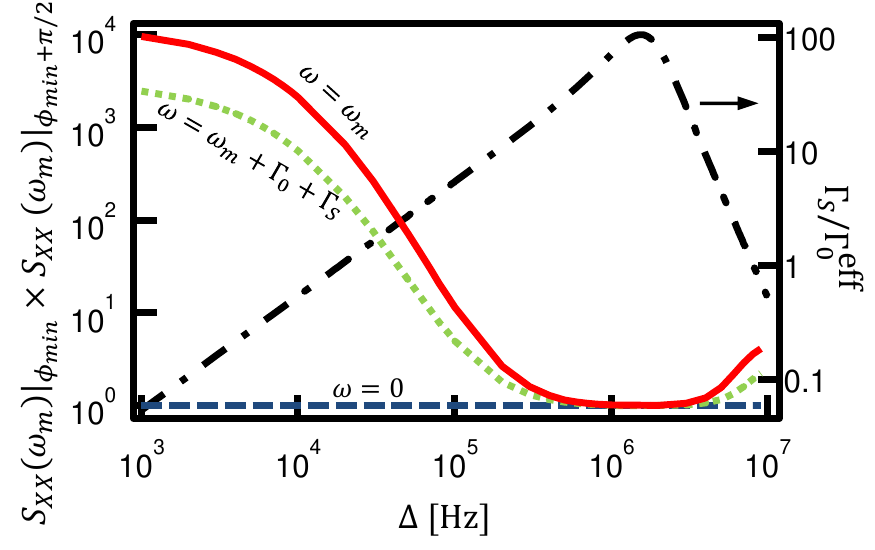}
	\caption{Approaching the Heisenberg uncertainty limit near the mechanical resonance.  The uncertainty product for the optical quadratures $S_{XX}(\omega)|_{\phi=\phi_{\mathrm{min}}} \times S_{XX}(\omega)|_{\phi=\phi_{\mathrm{min}}+\pi/2}$ is plotted at the mechanical resonance $\omega=\omega_m$ (solid red), slightly detuned from the mechanical resonance $\omega=\omega_m+\Gamma_S+\Gamma_0$ (dotted green), and far off the mechanical resonance $\omega\sim0$ (dashed blue).  $\Gamma_S$ is the optomechanical damping from the signal beam.  All three curves approach the Heisenberg uncertainty bound over a range of $\Delta$.  The ratio of signal beam induced optomechanical damping to intrinsic mechanical damping is also plotted (dot-dashed black).  The parameters used are: $g/2\pi=33$ Hz, $m=6.75\times 10^{-12}$~kg, $\omega_m^{\mathrm{eff}}/2\pi=1.5243$~MHz, $\Gamma_0^{\mathrm{eff}}/2\pi=2560$~Hz, $T^{\mathrm{eff}}=0$~K, $\epsilon_{\mathrm{ext}}=1$, $\kappa/2\pi=1.7$~MHz, $\kappa_R=\kappa$, $\bar{N}=1.1\times10^8$. }
	\label{fig:FigA2}
\end{suppfigure}

\section{Minimum Uncertainty States}

	The Heisenberg uncertainty limit for orthogonal quadratures of $S_{XX}$ can be written as $\sqrt{S_{XX}(\omega)|_{\phi=\phi_0}} \times \sqrt{S_{XX}(\omega)|_{\phi=\phi_0+\pi/2}}\geq 1$, for any $\phi_0$ and $\omega$.  It is interesting to note under what conditions this uncertainty product reaches the lower bound.  Optical loss and classical optical noise will add uncertainty to the optical field.  Thermal motion will induce excess noise on the light.  Eliminating all of these classical noise sources reduces the uncertainty product.  However, as shown in Fig.~S\ref{fig:FigA2}, the uncertainty product for the most deeply squeezed quadrature $\phi=\phi_{\mathrm{min}}$ with its orthogonal quadrature, only approaches 1 over some finite range of $\Delta$.  The curves of Fig.~S\ref{fig:FigA2} are plotted for an idealized system at zero temperature, but with finite damping, $\Gamma_0^{\mathrm{eff}}$, to a zero temperature bath, and no optical loss or classical noise.  For some values of $\Delta$, the ponderomotive effects of the light on the mechanical resonator can cause optimal correlations that allow the signal beam quadratures $\phi_{\mathrm{min}}$ and $\phi_{\mathrm{min}}+\pi/2$ to approach the Heisenberg uncertainty limit at frequencies near the mechanical resonance.  At frequencies far from the mechanical resonance, the mechanical excitation from both zero point motion and thermally driven motion is small, and the minimum uncertainty product can be obtained over a wide range of $\Delta$ even at finite temperature.

	With this understanding of relative level of ponderomotive squeezing and amplification, we are able to interpret the data in Fig.~3(b).  The agreement between theory and data indicates that the majority of the measured noise in the $\phi_{\mathrm{min}}+\pi/2$ quadrature is amplified quantum noise (with a small contribution from thermally induced mechanical motion at the level of $1/R$).  Thus we are able to infer a strong ponderomotive interaction strength from the large ($\sim 25$ dB) optical amplification rather than from the level of optical squeezing, where the latter is far more sensitive to optical loss.


\end{document}